\documentclass[9pt,twocolumn,twoside]{opticajnl}
\journal{opticajournal} 

\setboolean{shortarticle}{true}

\title{Recipe for near-unity efficient frequency conversion via transient molecular modulation}

\author[1]{Arturo Mena}
\author[1]{Pau Arcos}
\author[1]{María Sánchez-Hernández}
\author[2]{Amaia Berganza}
\author[1,3]{Joseba Zubia}
\author[1,3,4*]{David Novoa}

\affil[1]{Department of Communications Engineering, University of the Basque Country (EHU), Bilbao, 48013, Spain}
\affil[2]{Department of Applied Mathematics, University of the Basque Country (EHU), Bilbao, 48013, Spain}
\affil[3]{EHU Quantum Center, University of the Basque Country (EHU), Bilbao, Spain}
\affil[4]{IKERBASQUE, Basque Foundation for Science, Bilbao, 48009, Spain}

\affil[*]{david.novoa@ehu.eus}

\begin{abstract}
Frequency transduction technologies are an essential element for interfacing devices operating at distinct spectral regions. Specifically, the conversion of weak signals between different wavelengths is a fundamental requirement of quantum networks, where conversion efficiency is critical for preserving the statistical properties of light. In this regard, molecular modulation in gas-filled hollow-core fibers offers a promising approach for highly-tunable linear frequency conversion of classical and quantum light. However, recent experimental studies in this subject have struggled to overcome a saturation in conversion efficiency to about 70$~\%$, which, although high, might impair the faithful transfer of quantum properties upon conversion. In this work, we discuss a route to surpass this limitation that identifies the key factors to consider for obtaining quantum efficiencies exceeding 99\%. These results might pave the way to viable fiber-based quantum transduction schemes for next-generation quantum networks.
\end{abstract}

\setboolean{displaycopyright}{false} 

\begin{document}

\maketitle

Nonlinear frequency conversion is a key resource for extending the spectral reach of light sources and for interfacing systems operating at different wavelengths \cite{han2025}. Among the available approaches for frequency conversion, molecular modulation \cite{Harris1998} permits an exquisite control over the interacting fields, enabling e.g. the synthesis of arbitrary ultrafast light fields  \cite{Nazarkin1999}. While most early work on molecular modulation was performed in bulk materials such as gases or crystals, the advent of gas-filled hollow-core fibers (HCFs) \cite{benabid2002kagome} elevated this technique to new heights by lowering the threshold required to build sufficiently strong molecular coherence and enabling the ultralow-loss guidance of light fields over a broad bandwidth \cite{petrovich2025broadband}. As a result, molecular modulation in gas-filled HCFs \cite{arcos2024} constitutes a highly efficient and tunable method \cite{bauerschmidt2015broadband} to shift weak signals down to the single-photon level \cite{tyumenev2022}, to convert weak supercontinuum beams \cite{bauerschmidt2015broadband}, and to reach the deep-ultraviolet spectral domain \cite{mridha2019}.
\begin{figure}[h]
    \centering
    \includegraphics[width=\linewidth]{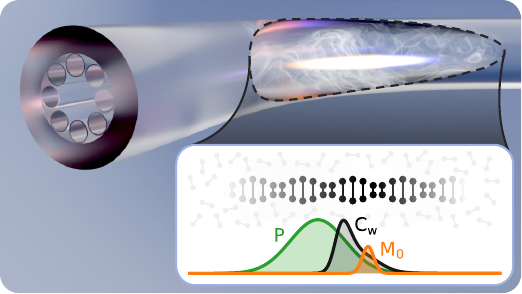}
    \caption{Illustration of molecular modulation in a gas-filled HCF. For efficient conversion of the M$_0$ pulse, time synchronization with the molecular coherence wave (C$_{\text{w}}$) is more effective than synchronizing with the pump (P) pulse. This condition is relaxed for pump durations longer than that of M$_0$.}
    \label{fig:intro_figure}
\end{figure}

\begin{figure*}[t!]
    \centering
    \includegraphics[width=\textwidth]{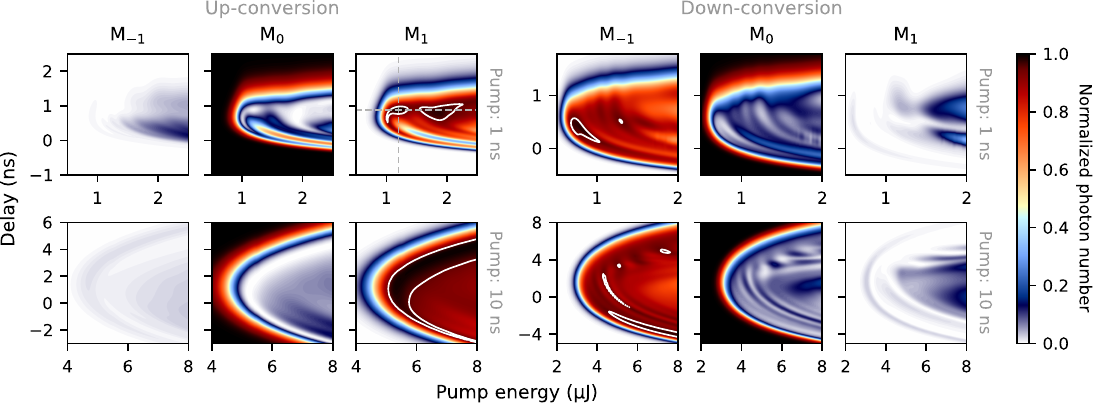}
    \caption{Normalized photon numbers as function of the pulse delay and pump energy at the output of a 70 cm-long hydrogen-filled HCF with core diameter of $\sim 20.7$ \textmu m and wall thickness of $\sim 120$ nm. The photon numbers are normalized with respect to the input M$_0$ photon population. Two pump durations are considered here: 1 ns (top row) and 10 ns (bottom row). The input M$_0$ pulse duration is 0.1 ns in all cases. White contour lines encircle the regions where the photon number population exceeds 95\% and the dashed lines in the M$_1$ subplot of up-conversion with a 1 ns-long pump indicate the case discussed at Fig.~\ref{fig:higheffcase}. }
    \label{fig:energy_delay_scan}
\end{figure*}
Molecular modulation relies on preparing a molecular vibrational coherence wave in advance by stimulated Raman scattering (SRS) \cite{Nazarkin1999, chen2024unified, Chen2026}. A pump pulse injected into the fiber generates a frequency down-shifted Stokes pulse from noise, and the beat note between the two drives a travelling coherence wave (see Fig.~\ref{fig:intro_figure}). Subsequently, a mixing pulse M$_0$ of frequency $\omega_{\text{M}_0}$, can scatter off this optical phonon wave if properly phase-matched and be selectively shifted by the Raman frequency $\Omega$ to either its anti-Stokes (M$_1$) or Stokes (M$_{-1}$) sideband, of frequencies $\omega_{\text{M}_{\pm 1}}=\omega_{\text{M}_0}\pm\Omega$. Since the molecular coherence is already established, this conversion has no threshold and can therefore operate on arbitrarily weak fields unlike conventional SRS \cite{tyumenev2022, mridha2019}. 

Among its several distinctive features, high conversion efficiency is particularly important in the quantum regime \cite{wang2022}. In the quantum description, molecular modulation acts as a frequency beam splitter \cite{gonzalez2025entanglement, wang2023quantum}, so any photons not transferred between modes can degrade the statistical properties of the light. A high internal efficiency is therefore a prerequisite for using this platform as a quantum frequency transducer. Recent experiments have shown the ability of molecular modulation to transfer biphoton correlations and photon-number statistics between frequencies \cite{tyumenev2022, mena2026}. However, a saturation in the exchange of photon-number population was always observed along propagation, making internal quantum efficiencies above roughly 70\% difficult to access \cite{bauerschmidt2015broadband, mridha2019, tyumenev2022, mena2026}. The origin of this saturation, and whether it is fundamental, has remained unclear. 

In this Letter, we use the semiclassical Maxwell-Bloch formalism of molecular modulation to trace the saturation to the interaction strength, which under phase-matched conditions is set by overlap integrals between the pulse envelopes and the coherence wave. As the pulses deplete, these overlaps vanish and the population exchange stalls. We show that the decisive condition for efficient conversion is not to synchronize the mixing pulse with the pump, but with the molecular coherence wave that truly drives the interaction, as represented in Fig.~\ref{fig:intro_figure}. Building on this, we give a simple recipe to overcome the saturation limit and yield internal conversion efficiencies exceeding 99\%.

The semiclassical description of molecular modulation extends the Maxwell-Bloch formalism of SRS \cite{raymer1990iii, arcos2024, chen2024unified} with $k$ additional field-envelope equations for the $\omega_k$ frequencies involved in the conversion that are kept below the Raman threshold. The field envelopes $E_k(z, \tau)$ depend on both the propagation distance $z$ along the fiber and the relative time $\tau$ measured in a reference frame co-propagating with the pulses. In the present work, we focus on the ``transient'' regime of vibrational SRS \cite{arcos2024, chen2024unified} with single-mode linearly co-polarized fields. In this regime, the pulse durations are comparable to the molecular dephasing time $T_2$, but are not strictly of the same order of magnitude. Therefore, considering only one converted sideband of the original mixing field, the equations driving the amplitude dynamics during molecular modulation are
\begin{align}
\frac{\partial E_{m+1}}{\partial z} &=-i\kappa_{2,m+1}\frac{\omega_{m+1}}{\omega_m} QE_me^{i(\beta_{m+1}-\beta_m)z},\label{eq:Em1-diff}\\
\frac{\partial E_m}{\partial z} &=-i\kappa_{2,m+1}Q^*E_{m+1} e^{-i(\beta_{m+1}-\beta_m)z}.
\label{eq:Em-diff}
\end{align}
Here, $\kappa_{2,m+1}$ is a coupling constant \cite{raymer1990iii} and $\beta_k$ denotes the modal propagation constant at frequency $\omega_k$. For up-conversion, the sideband subscripts take the form $\{m,m+1\}\rightarrow \{\text{M}_0, \text{M}_{1}\}$, while for down-conversion $\{m,m+1\}\rightarrow \{\text{M}_{-1}, \text{M}_0\}$. 

Meanwhile, the molecular coherence with amplitude $Q(z, \tau)$ is driven by the pump sidebands, which evolve according to the coupled SRS equations \cite{raymer1990iii} analogous to \eqref{eq:Em1-diff} and \eqref{eq:Em-diff}. The field envelopes of the pump sidebands are denoted as $E_\ell(z, \tau)$ and correspond to the frequencies $\omega_\ell=\omega_\text{P}+ \ell\Omega$, with $\omega_\text{P}$ being the pump frequency. Considering all of them, the equation describing the evolution of the molecular coherence amplitude $Q(z, \tau)$ is given as
\begin{equation}\label{eq:coherence}
\frac{\partial Q}{\partial \tau} 
 = -\frac{Q}{T_2} - \frac{i}{4} \sum_\ell\kappa_{1,\ell}E_\ell E_{\ell-1}^* e^{-i(\beta_\ell-\beta_{\ell-1})z},
\end{equation}
where $\kappa_{1,\ell}$ are coupling constants \cite{raymer1990iii} and we assumed that most of the molecules remain in the ground state. In most cases, the dominant bands driving the molecular coherence wave will be the pump (P) and its first Stokes sideband (S$_1$). Provided that energy conservation is satisfied, this yields to the phase-matching condition $\vartheta_{m+1,m}=(\beta_\text{P}-\beta_{\text{S}_1})-(\beta_{m+1} - \beta_m)=0$ for a given mixing conversion process $m\leftrightarrow m+1$. In gas-filled HCFs, the propagation constants are dependent on both the cross-sectional geometry of the fiber and on the properties of the filling gas. Hence, in practice, frequency up- or down conversion can be selected online by simply changing the gas pressure~\cite{mena2026}. 

To gain further understanding of the requirements for near-unity-efficient molecular modulation, let us consider a specific fiber geometry as an example. Concretely, without loss of generality, we consider the HCF fiber employed at Ref.~\cite{mena2026}, which is a kagome-style HCF with flat-to-flat core diameter of $\sim 20.7$~\textmu m and wall thickness of $\sim 120$~nm. In this setting, the fiber is pressurized with gaseous hydrogen. The vibrational Q(1) Raman transition of this gas enables a shift of $\Omega\sim 2\pi\times 125$~THz. As in Ref.~\cite{mena2026}, the pump wavelength will be $\sim 515$~nm and the mixing wavelength $\sim 585$~nm. Therefore, considering the capillary dispersion model \cite{finger2014accuracy}, the phase-matching pressures for up- and down-conversion are of $\sim 16.7$~bar and $\sim 28.1$~bar, respectively. However, instead of employing a fiber length of $\sim 3.7$~m, we will demonstrate that conversion efficiencies of over 95\% may be obtained even with a shorter length of e.g. 70~cm by adequate synchronization to the coherence wave. 

Fig.~\ref{fig:energy_delay_scan} presents mixing conversion simulations for different pump full width at half maximum durations, while the mixing duration is set to 0.1~ns in all of the cases. The simulations are carried out considering two Stokes and two anti-Stokes bands for both the pump and the mixing frequencies. Additionally, the input pulses are modeled as gaussian envelopes and the temporal walkoffs are neglected, as they are well below 1~ps/m at the working pressures. Because the mixing conversion dynamics are linear in intensity \cite{mridha2019}, for these examples we only consider guidance losses in the pump bands, which as can be seen in \eqref{eq:coherence} govern the evolution landscape of the coherence wave.

In Fig.~\ref{fig:energy_delay_scan} we can observe that for pump durations closer to the duration of the mixing pulse, like the cases with a 1 ns-long pump, delaying the mixing pulse to the trailing edge of the pump pulse (positive delay values) brings the mixing to the temporal region where the coherence wave dominates. This results in an increase of the population in the converted band, even at pump energies with which no population transfer would be observed at zero delay, where the input mixing pulse is perfectly aligned with the input pump at $\tau=0$. In addition to this, for the 1 ns-long pump case, the parametric scan becomes more structured in comparison to the cases considering a pump duration of 10 ns, where the scan becomes smoother. This implies that, although the energies required with shorter pump pulses are lower, the conversion process becomes more sensitive to experimental energy fluctuations and requires a more precise delay adjustment. Therefore, an adequate balance between pulse durations and pump energies in the fiber core is fundamental for extending the high-efficiency regions in the parameter space. Furthermore, it is also interesting to operate in the regions corresponding to lower pump energies injected into the fiber core in order to minimize the crosstalk with unwanted mixing bands during the conversion, which limit the maximum photon population transferred to the desired sideband. This crosstalk may be fostered by e.g. the presence of large coherence wave amplitudes, capable of initiating high-dephasing mixing transitions, or by the presence of pump sideband pairs other than the P-S$_1$ pair that introduce additional components to the coherence. 

The maximum population values in the converted bands at Fig.~\ref{fig:energy_delay_scan} for the cases with a 1~ns pump are 99.1\% for up-conversion with a delay of 0.88~ns and a pump energy of 1.2~\textmu J, while for down-conversion the value is 97.7\% at a delay of 0.38~ns and a pump energy of 0.8~\textmu J. With a 10 ns-long pump, the maximum is 99.2\% with a delay of 4.3~ns and a pump energy of 8~\textmu J for up-conversion, and 96.9\% for down-conversion with a $-4.1$~ns delay and 8~\textmu J of pump energy. 

\eqref{eq:Em1-diff} and \eqref{eq:Em-diff} indicate that the rate of amplitude transfer between bands is modulated by the products $Q(z, \tau) E_m(z, \tau)$ and $Q^*(z, \tau) E_{m+1}(z, \tau)$, aside from being also conditioned by the exponential dephasing terms that give the phase-matching conditions that we previously discussed. Therefore, to maximize the photon population transfer between mixing bands in a phase-matched setting, these products need to be adjusted both along $z$ and $\tau$, which is done by e.g. synchronizing the input mixing beam with the molecular coherence wave as we saw in Fig.~\ref{fig:energy_delay_scan}. This condition also appears when considering the analytical solutions of the Maxwell-Bloch equations with negligible mixing depletion, which are analogous to those describing Stokes generation with negligible pump depletion \cite{raymer1990iii} and yield overlap integrals between envelopes.

\begin{figure}[t]
    \centering
    \includegraphics[width=\linewidth]{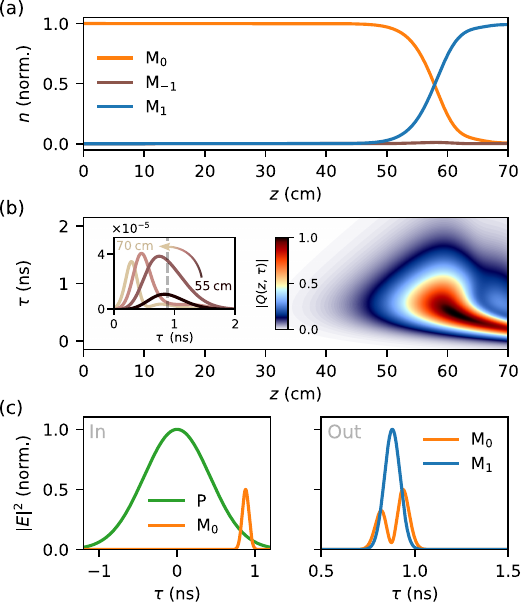}
    \caption{Dynamics of a frequency up-conversion process with $99.1$\% quantum efficiency corresponding to the case in Fig.~\ref{fig:energy_delay_scan} with a 1 ns-long pump. (a) Evolution of the photon-number populations normalized with respect to the input M$_0$ photon number. (b) Spatio-temporal dynamics of the molecular coherence wave along the fiber. The amplitude is normalized to the maximum, while the inset shows the unnormalized envelopes at four slices in $z$ with a dashed line indicating the delay of the input M$_0$ pulse. (c) Relevant envelopes at the HCF input and output. The envelopes are normalized to their own maximum, while the M$_0$ envelopes are additionally scaled by $\times 0.5$ to indicate a lower intensity than the neighboring envelope.}
    \label{fig:higheffcase}
\end{figure}

This also explains the origin of the saturation in conversion efficiency observed in experiments even with phase matching. During the dynamics, the envelopes may for example deplete too soon or desynchronize with the coherence wave, making the interaction strength given by the overlaps vanish. However, maximizing the interaction along the whole fiber length can also be detrimental for the output efficiency. For example, if the fiber length is too large or the coherence wave is too intense, when the mixing photon population is fully transferred into the converted band, the photons in this band may start being converted back into the mixing band, giving rise to oscillations in the photon number population. Therefore, the key point here to both mitigate the saturation effects and maximize the output conversion efficiency is not the synchronization of the mixing pulse with the pump, nor with the coherence wave at the relative time $\tau$ in which the maximum coherence amplitude is found, but to adequately synchronize the mixing with the whole evolution of the molecular coherence wave driving the interaction. This is representative of the importance of the time domain in this phenomenology, as pointed out in other works \cite{chen2024unified, Chen2026}. 

\begin{figure}[t]
    \centering
    \includegraphics[width=\linewidth]{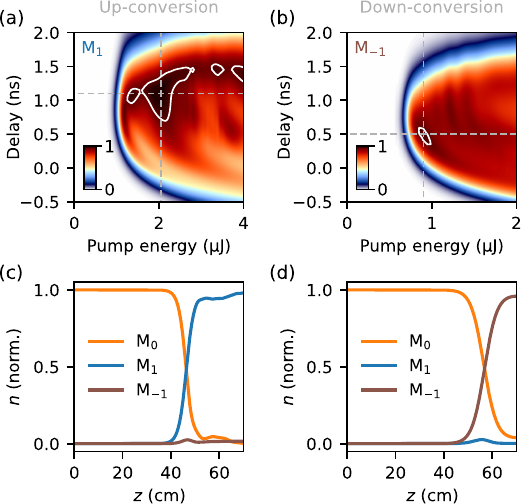}
    \caption{Optimization of molecular modulation with the parameters of Ref.~\cite{mena2026} but only with 70~cm of HCF. (a) Photon number in the M$_1$ band after up-conversion normalized with respect to the input M$_0$ value as a function of the pump energy and the delay. (b) Same as (a) but for the M$_{-1}$ band during down-conversion. (c) and (d) show the normalized photon number dynamics as a function of $z$ in the maximum efficiency cases of (a) and (b) respectively, which are marked with dashed lines.}
    \label{fig:expcaseoptimized}
\end{figure}

We can exemplify this discussion by considering the case of 99.1\% efficiency in Fig.~\ref{fig:energy_delay_scan}, where the mixing delay was set to 0.88~ns. In this regard, Fig.~\ref{fig:higheffcase} presents the dynamics of the photon numbers (Fig.~\ref{fig:higheffcase}a) and the coherence wave amplitude $|Q(z, \tau)|$ (Fig.~\ref{fig:higheffcase}b) as light propagates through the fiber. The relevant pulse envelopes at the input and output of the fiber are shown at Fig.~\ref{fig:higheffcase}c. In Fig.~\ref{fig:higheffcase}b we can see that the coherence wave rises at relative pulse times $\tau$ in the region where the mixing pulse is found (see Fig.~\ref{fig:higheffcase}c), initiating the conversion and allowing the efficient transfer of photon population to the up-converted band. As the population transfer reaches its maximum, the peak of the coherence wave is conveniently displaced outside of the temporal reach of the mixing envelope (see inset of Fig.~\ref{fig:higheffcase}b), ceasing the interaction and avoiding Rabi-like oscillations in the photon number populations that would impair the output efficiency. This deeper understanding of the dynamics can be applied to current experimental setups in the literature. For instance, for the conditions at Ref.~\cite{mena2026} with 1.42~ns and 0.3~ns-long pump and mixing pulses, the simulations in Fig.~\ref{fig:expcaseoptimized} show that, with for example only 70~cm of fiber, internal quantum efficiencies of 98.1\% and 96\% can readily be obtained for up- and down-conversion respectively.

In conclusion, we have explored the origin of the saturation in frequency conversion efficiency during phase-matched molecular modulation in gas-filled HCFs. An analysis on the semiclassical Maxwell-Bloch theory in the transient regime \cite{raymer1990iii} pointed out to the the importance of proper synchronization of the mixing pulses to be converted with the whole spatio-temporal evolution of the molecular coherence wave, which strongly modulates the interaction. By adequately controlling the overall dynamics  through a judicious choice of pump energy, mixing pulse delay and pump-mixing duration ratio, the saturation limit can be surpassed, enabling seamless access to near-unity efficiencies. Our numerical modelling indicates that the best approach is to select pump durations longer than the mixing pulse, to allow the envelope of the molecular coherence wave to be of a duration comparable to that of the mixing. This also enhances the stability of the conversion process, as it extends the high-efficiency regions in the parameter space, relaxing the demands for pump energy stability and mixing synchronization, which is interesting for compensating the jitter of photon sources like quantum dot emitters \cite{buckley2012engineered}. In the future, the combination of this approach for efficient molecular modulation with the latest ultralow-loss HCF designs \cite{petrovich2025broadband, sabbah2025} may unlock the design of experimental setups with both high internal and external (i.e. including optical loss) quantum transduction efficiencies. This can be critical for applications that require the preservation of quantum properties such as entanglement between photonic frequency subspaces \cite{knorzer2026}, as the metrics employed in these scenarios are closely related to the photon populations \cite{gonzalez2025entanglement}.

\begin{backmatter}
\bmsection{Funding} This work was supported by the grants PID2024-158310NB-I00, PID2024-155582OB-C31, CNS2025-166590 and PRE2022-102843, funded by MICIU/AEI/10.13039/501100011033, by ERDF, EU, and "ESF+", by Gobierno Vasco/Eusko Jaurlaritza under Grant IT1829-26, ELKARTEK Smart\textmu S 2025/00058 and Newhegaz KK 2025/00074, by “Translight” Initiative of EHU, and by the IKUR Strategy of the Department of Education of the Basque Government under Grant IKUR-IKA-23/03. MS-H acknowledges support from the predoctoral grant “Formación de Profesorado Universitario” FPU22/01451 from the Spanish Ministry of Science, Innovation and Universities (MICIU).

\bmsection{Disclosures} The authors declare no conflicts of interest.

\bmsection{Data availability} Data underlying the results presented in this paper may be obtained from the authors upon reasonable request.

\end{backmatter}

\bibliography{bibliography}

@article{han2025,
  title={Recent Advances in $\chi^{(2)}$-Based Nonlinear Frequency Conversion of Fiber Lasers},
  author={Han, Bing and Tao, Yiming and Deng, Wei and Chen, Lu and Hu, Bo and Cheng, Qian and Qiu, Yihan and Fan, Mengqiu and Liang, Houkun and Feng, Ting and others},
  journal={Advanced Photonics Research},
  volume={6},
  number={12},
  pages={2500012},
  year={2025},
  publisher={Wiley Online Library}
}

@article{arcos2024,
year = {2024},
month = {aug},
publisher = {EDP Sciences, IOP Publishing and Società Italiana di Fisica},
volume = {147},
number = {4},
pages = {45001},
author = {Arcos, P. and Mena, A. and Sánchez-Hernández, M. and Arrospide, E. and Aldabaldetreku, G. and Illarramendi, M. A. and Zubia, J. and Novoa, D.},
title = {{Narrowband stimulated Raman scattering and molecular modulation in anti-resonant hollow-core fibres}},
journal = {EPL}
}

@article{bauerschmidt2015broadband,
  title={{Broadband-tunable LP01 mode frequency shifting by Raman coherence waves in a H2-filled hollow-core photonic crystal fiber}},
  author={Bauerschmidt, ST and Novoa, D and Abdolvand, A and Russell, P St J},
  journal={Optica},
  volume={2},
  number={6},
  pages={536--539},
  year={2015},
  publisher={Optical Society of America}
}

@article{tyumenev2022,
author = {R. Tyumenev  and J. Hammer  and N. Y. Joly  and P. St. J. Russell  and D. Novoa },
title = {{Tunable and state-preserving frequency conversion of single photons in hydrogen}},
journal = {Science},
volume = {376},
number = {6593},
pages = {621-624},
year = {2022}
}

@article{mridha2019,
author = {Manoj K. Mridha and David Novoa and Pooria Hosseini and Philip St.J. Russell},
journal = {Optica},
number = {6},
pages = {731--734},
publisher = {Optica Publishing Group},
title = {Thresholdless deep and vacuum ultraviolet Raman frequency conversion in hydrogen-filled photonic crystal fiber},
volume = {6},
year = {2019},
}

@article{wang2022,
  title={{Quantum capacities of transducers}},
  author={Wang, Chiao-Hsuan and Li, Fangxin and Jiang, Liang},
  journal={Nature Communications},
  volume={13},
  number={1},
  pages={6698},
  year={2022},
  publisher={Nature Publishing Group UK London}
}

@article{gonzalez2025entanglement,
  title={{Entanglement transfer during quantum frequency conversion in gas-filled hollow-core fibers}},
  author={Gonzalez-Raya, Tasio and Mena, Arturo and Lazo, Miriam and Leggio, Luca and Novoa, David and Sanz, Mikel},
  journal={APL Photonics},
  volume={10},
  number={4},
  year={2025},
  publisher={AIP Publishing}
}

@article{wang2023quantum,
  title={{Quantum entanglement between signal and frequency-up-converted idler photons}},
  author={Wang, Jiaxuan and Sokolov, Alexei V and Agarwal, Girish S},
  journal={Physical Review A},
  volume={108},
  number={6},
  pages={063706},
  year={2023},
  publisher={APS}
}

@article{mena2026,
      title={Preservation of photon statistics upon selective molecular modulation in hollow-core fibers}, 
      author={Arturo Mena and Pau Arcos and María Sánchez-Hernández and Benoit Debord and Frédéric Gérôme and Fetah Benabid and Nicolas Y. Joly and Oskar Arrizabalaga and Joseba Zubia and David Novoa},
      year={2026},
      eprint={2610.19964},
      archivePrefix={arXiv},
      journal={arXiv:2610.19964},
      url={https://arxiv.org/abs/2610.19964}, 
}

@article{raymer1990iii,
  title={{III The quantum coherence properties of stimulated Raman scattering}},
  author={Raymer, MG and Walmsley, IA},
  journal={Progress in Optics},
  volume={28},
  pages={181--270},
  year={1990},
  publisher={Elsevier}
}

@article{chen2024unified,
  title={{Unified and vector theory of Raman scattering in gas-filled hollow-core fiber across temporal regimes}},
  author={Chen, Yi-Hao and Wise, Frank},
  journal={APL Photonics},
  volume={9},
  number={3},
  pages = {030902},
  year={2024},
  publisher={AIP Publishing}
}

@article{Chen2026,
author = {Yi-Hao Chen and Wenchao Wang and Shaodong Hou and Jose Enrique Antonio-Lopez and Rodrigo Amezcua-Correa and Chris Xu and Frank Wise},
journal = {Optica},
number = {11},
pages = {2217--2227},
publisher = {Optica Publishing Group},
title = {Highly efficient perturbative Raman shifting by engineering the nonlinear temporal response},
volume = {13},
year = {2026}
}

@article{Nazarkin1999,
    author = {A. Nazarkin and G. Korn and M. Wittmann and T. Elsaesser},
    title = {Generation of Multiple Phase-Locked {Stokes} and Anti-{Stokes} Components in an Impulsively Excited {Raman} Medium},
    journal = {Phys. Rev. Lett.},
    volume = {83},
    pages = {2560},
    year = {1999},
    doi = {10.1103/PhysRevLett.83.2560}
}

@article{Harris1998,
    author = {S. E. Harris and A. V. Sokolov},
    title = {Subfemtosecond Pulse Generation by Molecular Modulation},
    journal = {Phys. Rev. Lett.},
    volume = {81},
    pages = {2894},
    year = {1998},
    doi = {10.1103/PhysRevLett.81.2894}
}

@article{finger2014accuracy,
  title={Accuracy of the capillary approximation for gas-filled kagom{\'e}-style photonic crystal fibers},
  author={Finger, Martin A and Joly, Nicolas Y and Weiss, Thomas and Russell, P St J},
  journal={Optics Letters},
  volume={39},
  number={4},
  pages={821--824},
  year={2014},
  publisher={Optical Society of America}
}

@article{sabbah2025,
  title={{Resonance-free deep ultraviolet to near infrared supercontinuum generation in a hollow-core antiresonant fibre}},
  author={Sabbah, Mohammed and Mears, Robbie and Harrington, Kerrianne and Wadsworth, William J and Stone, James M and Birks, Tim A and Travers, John C},
  journal={Journal of Physics: Photonics},
  volume={7},
  number={2},
  pages={025019},
  year={2025},
  publisher={IOP Publishing}
}

@article{petrovich2025broadband,
  title={{Broadband optical fibre with an attenuation lower than 0.1 decibel per kilometre}},
  author={Petrovich, Marco and Numkam Fokoua, Eric and Chen, Yong and Sakr, Hesham and Adamu, Abubakar Isa and Hassan, Rosdi and Wu, Dong and Fatobene Ando, Ron and Papadimopoulos, Athanasios and Sandoghchi, Seyed Reza and others},
  journal={Nature Photonics},
  volume={19},
  number={11},
  pages={1203--1208},
  year={2025},
  publisher={Nature Publishing Group UK London}
}

@article{buckley2012engineered,
  title={Engineered quantum dot single-photon sources},
  author={Buckley, Sonia and Rivoire, Kelley and Vu{\v{c}}kovi{\'c}, Jelena},
  journal={Rep. Prog. Phys.},
  volume={75},
  number={12},
  pages={126503},
  year={2012},
  publisher={IOP Publishing}
}

@article{knorzer2026,
  title={{Distributed quantum information processing: a review of recent progress}},
  author={Kn{\"o}rzer, Johannes and Liu, Xiaoyu and Schiffer, Benjamin F and Tura, Jordi},
  journal={Rep. Prog. Phys.},
  volume={89},
  number={7},
  pages={074401},
  year={2026},
  publisher={IOP Publishing}
}

@article{benabid2002kagome,
  title={Stimulated Raman scattering in hydrogen-filled hollow-core photonic crystal fiber},
  author={Benabid, Fetah and Knight, Jonathan and Antonopoulos, G and St.J. Russell, Philip},
  journal={Science},
  volume={298},
  number={5592},
  pages={399-402},
  year={2002},
  publisher={Science}
}

\end{document}